\pdfoutput=1 
\documentclass[aps,prd,twocolumn,superscriptaddress,tightelines,nofootinbib]{revtex4}
\usepackage{graphicx}
\usepackage{epsfig}
\usepackage{bm}
\usepackage{latexsym,amssymb,amsmath,amsfonts,amssymb,txfonts,pxfonts,wasysym,float}
\usepackage{mathrsfs}
\usepackage{color}
\usepackage{lettrine}
\usepackage{lipsum}
\usepackage{enumitem}

\usepackage[usenames,dvipsnames]{xcolor}
\definecolor{orange}{cmyk}{0,0.5,1,0}
\definecolor{rossoCP3}{cmyk}{0,.88,.77,.40}
\definecolor{graa}{rgb}{0.8,0.8,0.8}
\definecolor{blaa}{rgb}{0.2,0.2,0.6}

\begin{document}

\title{\color{rossoCP3} Upper limit on the fraction of alien civilizations that develop communication technology}

\author{Luis A. Anchordoqui}
\affiliation{Department of Physics \& Astronomy,  Lehman College, City University of
  New York, NY 10468, USA}
\affiliation{Department of Physics,
 Graduate Center, City University
  of New York,  NY 10016, USA}
\affiliation{Department of Astrophysics,
 American Museum of Natural History, NY
 10024, USA}

\author{Susanna M. Weber}
\affiliation{Mamaroneck High School, 1000 Boston Post Rd., Mamaroneck, NY 10543, USA}
\date{August 2019}

\begin{abstract}
  \noindent We re-examine the likelihood for alien civilizations to develop communication technology on the basis of the general assumption that life elsewhere could have a non-carbon chemical foundation. We particularized the discussion to a complex silicon-based biochemistry in a nitrogen solvent, and elaborate on the environment in which such a chemistry is feasible, and if so, on what scales. More concretely, we determine the region outside the habitable zone where such organisms can grow and flourish and after that we study how our findings impact the recently derived upper limit on the fraction of living intelligent species that develop communication technology $\langle \xi_{\rm biotec} \rangle$.  We also compare this new restriction on $\langle \xi_{\rm biotec} \rangle$ with that resulting from the extension of the habitable zone to accommodate subsurface exolife, originating in planets with subsurface (water) oceans.

\end{abstract}

\maketitle

\section{Introduction}

\lettrine[lraise=0.1, nindent=0em, slope=-.5em]{T}{he}
 day on which the Earth began to move is a seminal moment in world history. In the XVI century, Copernicus' theory of heliocentrism transformed a millennia-old worldview with the shocking revolutionary idea that we humans are not at the center of the universe.  
We may be again at the verge of an exciting moment in the history of mankind. In the last few decades, the so-called ``exoplanet'' revolution has shown us that the universe is awash in alien worlds and that we humans may not be alone. The heedful quest for ``exolife'' has been dubbed the {\it cosmic modesty conjecture}: ``The richness of the universe teaches us modesty and guides us to search for both primitive and intelligent forms of life elsewhere without prejudice''~\cite{Loeb:2017cm}.

In its most restrictive incarnation the cosmic modesty conjecture calls for carbon-based organisms, operating in a water-based medium, with higher forms (perhaps) metabolizing oxygen. All forms of life on Earth share this same basic biochemistry. Indeed, the concept of life based around anything other than carbon certainly seems outlandish at first. However, our present knowledge of physics does not guarantee this fact, rather falling into what Sagan referred to as ``carbon chauvinism''~\cite{Sagan}. In this work
we re-examine the likelihood for alien civilizations to develop communication technology.  We adopt an extreme viewpoint of the cosmic modesty conjecture and argue that life is a sort-of ``nanotechnology phenomenon'' of ``molecular automaton'' style, and that it is the liquid nature where life evolves what determines the biochemistry of non-terrestrial life (rather than a certain biochemistry being required for life's existence).

The number of intelligent civilizations in our galaxy at any given time capable of releasing detectable signals of their existence into space can be cast in a quite simple functional form, 
\begin{equation}
 N = \Gamma_\star \ f_{\rm p}  \ n_{\rm e}  \ f_{\ell} \ f_{\rm i} \ f_{\rm c} \ L_\tau \,,
\label{Drake}
\end{equation}
where $\Gamma_\star$ is the average rate of star formation, $f_{\rm p}$ is the fraction of stars with planetary systems, $n_{\rm e}$ is the number of planets (per solar system) with a long-lasting ($\sim 4~{\rm Gyr}$)   ecoshell, $f_\ell$ is the fraction of suitable planets on which life actually appears, $f_{\rm i}$ is the fraction of living species that develop intelligence, $f_{\rm c}$ is the fraction of intelligent species with communications technology, and $L_\tau$ is the length of time such civilizations release detectable signals into space (i.e. the lifetime of the {\it communicative phase})~\cite{Drake}.

A more compact form of (\ref{Drake}) can be obtained by considering separately  its astrophysical and biotechnological factors
\begin{equation}
N = \langle \zeta_{\rm astro} \rangle \, \langle \xi_{\rm biotec} \rangle \, L_\tau \, ,
\label{Dshort}
\end{equation}
where $\langle \zeta_{\rm astro} \rangle = \Gamma_\star \, f_{\rm p} \, n_{\rm e}$ represents the production rate of habitable planets with long-lasting ecoshell (determined through astrophysics) and $\langle \xi_{\rm biotec} \rangle = f_\ell \, f_{\rm i} \, f_{\rm c}$ represents the product of all chemical, biological and technological factors leading to the development of a technological civilization~\cite{Prantzos:2013hg}. $\langle \cdots \rangle$ indicates average over all the multiple manners civilizations can arise, grow, and develop such technology, starting at any time since the formation of our Galaxy in any location inside it. This averaging procedure must be regarded as a crude approximation because the characteristics of the initial conditions in a planet and its surroundings may affect $f_\ell$, $f_{\rm i}$, and $f_{\rm c}$  with high complexity.  In a recent study we estimated the production rate of exoplanets where carbon-based organisms operating in a water-based medium can flourish  and the rate of planetary catastrophes which  could threaten the evolution of life on the surface of these worlds~\cite{Anchordoqui:2017eog}.
Armed with these estimates we used our current measurement of $N = 0$ to set an upper limit on $\langle \xi_{\rm biotec} \rangle$. In this work we extend our study by considering  that life elsewhere could have a non-carbon chemical foundation.

The layout of the paper is as follows. We begin in Sec.~\ref{sec:2} with an overview of the circumstellar habitable zone. In Sec.~\ref{sec:3} we explore wether life may develop  outside the habitable zone. 
In Sec.~\ref{sec:4} we combine the odds of finding exolife different from us with the non observation of artificial signals from beyond Earth to revise the upper limit on the average fraction of living intelligent species that develop communication technology. We conclude in Sec.~\ref{sec:5} with some implications of our results.

\section{The Habitable Zone}
\label{sec:2}

\begin{table*}
  \caption{Characteristics of some planets in the solar system. $\mathscr{F}_{r_p}$ is the solar constant at a distance $r_p$ from the Sun, $\alpha_p$ is the planetary albedo, $T_p$ is the emission temperature of the planer estimated using (\ref{Tp}), $T_{p_{\rm measured}}$ is the measured emission temperature, and $T_{p_{\rm surface}}$ is the global mean surface temperature of the planet. The rotation period $\tau$ is given in Earth days. \label{tabla1}}
 \begin{tabular}{cccccccc}
\hline
\hline
~~~~~planet~~~~~ &~~~~~$r_p$ ($10^9$~m)~~~~~&~~~~~$\mathscr{F}_{r_p}$ (W m$^{-2}$)~~~~~&~~~~~$\alpha_p$~~~~~&~~~~~$T_p$ (K)~~~~~&~~~~~$T_{p_{\rm measured}}$ (K)~~~~~&~~~~~$T_{p_{\rm surface}}$ (K)~~~~~& ~~~~~$\tau$ (Earth days)~~~~~ \\
\hline
Venus & 108 & 2632 & 0.77 & 227 & 230 & 760 & 243 \\
Earth & 150 & 1367 & 0.30 & 255 & 250 & 288 & 1.00 \\
Mars & 228 & 589 & 0.24 & 211 & 220 & 230 & 1.03 \\
Jupiter & 780 & 51 & 0.51 & 103 & 130 & 134 & 0.41 \\
\hline
\hline
\end{tabular}
\end{table*}

Life needs of stars for at least two reasons: 
\begin{itemize}[noitemsep,topsep=0pt]
\item stars
are required to synthesize heavy elements (such as carbon, oxygen,
$\cdots$, iron) out of which rocky planets and the molecules of life
are made;
\item stars maintain a source of heat to
power the chemistry of life on the surface of their planets. 
\end{itemize}
For example, the Earth receives almost all of its energy from the Sun. We can think of the Sun as a blackbody radiator at temperature $T_\odot \simeq 5,777~{\rm K}$. The Stefan-Boltzmann law gives us a way of calculating the power radiated per unit area,
\begin{equation}
{\rm Sun's \ emitted \ radiation \ per \  unit \ area} = \sigma T_\odot^4,
\label{uno}
\end{equation}
where $\sigma = 5.67 \times 10^{-8}~{\rm W m^{-2} K^{-4}}$. At the present time in its evolution the Sun emits energy at a rate of
\begin{eqnarray}
  {\rm Solar \ luminosity} \equiv  L_\odot & = & 4 \pi \ R_\odot^2 \ \sigma \ T_\odot^4 \nonumber \\
  & \simeq & 3.87 \times 10^{26}~{\rm W} \, ,
\label{dos}
\end{eqnarray}
where we have assumed that the Sun is a sphere of radius $R_\odot = 6.95 \times 10^8~{\rm m}$.

The flux of solar energy at the Earth, generally refer to as the ``solar constant'', depends on the distance of the Earth from the Sun, $r_\oplus$, and is given by the inverse square law
\begin{equation}
{\rm  Solar \ constant} \equiv  \mathscr{F}_{r_\oplus} = \frac{L_\odot}{4\pi r_\oplus^2} \, .
\label{tres}
\end{equation}
Obviously, due to variations in the orbit of the Earth the solar constant is not actually  constant. The estimate  set out in Table~\ref{tabla1},  $\mathscr{F}_{r_\oplus} = 1,367~{\rm W m}^{-2}$, corresponds to the average value which results from the average Earth-Sun distance of $r_\oplus = 1.5 \times 10^{11}~{\rm m}$.

\begin{table}
 \caption{Albedos for different surfaces. \label{tabla2}}
  \begin{tabular}{lc}
  \hline \hline
type of surface~~~~~~~~ & ~~~~~~~albedo (\%)~~~~~~~ \\
\hline
    ocean & 2 - 10 \\
  forest & 6 - 18\\
  cities & 14 - 18 \\
  grass & 7 - 25 \\
  soil & 10 - 20 \\
  grassland & 16 - 20 \\
  desert (sand) & 35 - 45 \\
  ice & 20 -70 \\
  cloud (thin, thick stratus) & 3, 60 -70 \\
  snow (old) & 40 -60 \\
  snow (fresh) & 75 - 95 \\
  \hline
  \hline
\end{tabular}
\end{table}

 Since water is essential
for life as we know it, the search for biosignature gases naturally
focuses on planets located in the {\it habitable zone} of their host stars,
which is defined as the orbital range around the star within which surface liquid
water could be sustained.  Each star is surrounded by a {\it habitable zone}.

To estimate the orbital range of the habitable zone in the solar system we 
consider the energy balance of any planet intercepting the solar energy flux and radiating energy away. Suppose that a planet of radius $R_p$ absorbs all incident light (100\%) from the Sun.  The planet blocks out an area of $\pi R_p^2$, and so this is the area we will use to find the power that is absorbed by the planet. Note that we do not use the whole surface area of the spherical planet, since the whole planet does not intercept the sunlight all at the same time. We also do not use half of the whole surface area of the spherical planet, since even though half of the planet is in ``daylight'', the areas that are not perpendicular to the rays of sunlight should not be given as much importance as the areas that are perpendicular to the rays of sunlight. We need to use the two dimensional area blocked out by the spherical planet, and since a sphere always casts a circular shadow, we need only find the area of a circle of radius $R_p$.  The power input to the planet is given by
\begin{equation}
{\rm Solar \ power \ incident \ on \ planet} = \mathscr{F}_{r_p} \ \pi  \ R_p^2 \, ,
\end{equation}
and so for planet Earth we have
\begin{eqnarray}
{\rm Solar \ power \ incident\  on \  Earth} &  =  & \mathscr{F}_{r_\oplus} \ \pi  \ R_\oplus^2 \nonumber \\ & \simeq & 1.74 \times 10^{17}~{\rm W} \, .
\end{eqnarray}
However, not all of this radiation is absorbed by the planet, but rather a significant fraction is reflected. The ratio of reflected to incident solar energy is called the albedo, $\alpha_p$, which depends on the nature of the reflecting surface  (it is large for clouds and light surfaces such as deserts, snow, and ice), see Table~\ref{tabla2}. Under the present terrestrial conditions of cloudiness and snow and ice cover, on average a fraction $\alpha_\oplus \sim 0.30$ of the incoming solar radiation at the Earth is reflected back to space. The solar radiation absorbed by a planet is then
\begin{equation}
\mathscr{F}_{p,{\rm abs}} = (1 -\alpha_p) \ \mathscr{F}_{r_p} \ \pi R_p^2 \, .
\label{seis}
\end{equation}
Estimates of exoplanet's albedos are given in~\cite{DelGenio}. In equilibrium, the total flux radiated by the planet into space must balance the radiation it absorbes. Now, we assume that the planets emits in all directions like a blackbody of uniform temperature $T_p$ (known as the ``effective planetary temperature'', or ``emission temperature'') the Stefan-Boltzmann law gives:
\begin{equation}
{\rm Planet's\ emitted \ radiation \ per \  unit \ area} = \sigma T_p^4,
\end{equation}  
and so the total radiation emitted by the planet is
\begin{equation}
  {\rm Planet's\ emitted \ radiation} \equiv \mathscr{F}_{p,{\rm em}} = 4 \pi\  R_P^2 \ \sigma \ T_p^4 \,.
  \label{ocho}
\end{equation}
Note that (\ref{ocho}) predicts the temperature one would infer by looking at the planet if a black body curve were fitted to the measured spectrum of outgoing radiation, and therefore this can be taken as the definition of the emission temperature. We can  now equate (\ref{seis}) and (\ref{ocho}) to obtain  
\begin{equation}
  T_p = \left[ \frac{\mathscr{F}_{r_p} (1 -\alpha_p)}{4\sigma} \right]^{1/4} \, .
\label{Tp}
\end{equation}
It is noteworthy that the radius of the Earth has cancelled out. This means that  $T_p$ depends only on the planetary albedo and the distance of the planet from the Sun. Substituting for the values given in Table~\ref{tabla1} we find that the Earth has an emission temperature of $T_\oplus \sim 255~{\rm K}$.

Using (\ref{dos}) and (\ref{tres}) we  estimate the flux of solar energy for any planet in the solar system,
\begin{equation}
\mathscr{F}_{r_p} = \frac{R_\odot^2 \sigma T_\odot^4}{r^2_p }\,.
\label{diez}
\end{equation}
Substituting (\ref{diez}) into (\ref{Tp}) and rearranging we obtain 
\begin{equation}
  r_p = \frac{R_\odot \, T_\odot^2 \, \sqrt{1 - \alpha_p} }{2 \ T_p^2} \, .
  \label{once}
\end{equation}
Using (\ref{once}) it is straightforward to estimate the range of distances that determine the habitable zone by requiring $273  < T_p/{\rm K} < 373$. 

In Table~\ref{tabla1} we show several  parameters for some of the planets in the solar system and we compare the approximate measured values with those computed using (\ref{Tp}). In general, there is a good agreement.  Jupiter, however, is an exception
due to the fact that about one-half of the energy input comes from the gravitational collapse of the planet.

A point worth noting at this juncture is that the temperature of a planet as derived in (\ref{Tp}) has small corrections due to atmospheric effects. More concretely, the so-called ``natural greenhouse effect'' is the process by which radiation from a planet's atmosphere warms the planet's surface to a temperature above what it would be without this atmosphere. Radiatively active gases (i.e., greenhouse gases) in a planet's atmosphere radiate energy in all directions. Obviously, such a  greenhouse effect is not the same on all planets, and differs dramatically based on the thickness and composition of the atmosphere. Three planets that show how dramatically the conditions of a planet can change with the different levels of the greenhouse effect are Venus, Earth, and Mars. For example, $T_\oplus$ is nearly 40~K cooler than the globally averaged observed surface temperature. Without the greenhouse effect, the temperature on Earth would be too cold for life.

There are four main naturally occurring gases that are responsible for the greenhouse effect: water vapor, carbon dioxide, methane and nitrous oxide. Of these gases, water vapor has the largest effect. Once these gases absorb energy, the gas particles begin to vibrate and they radiate energy in all directions, including approximately 30\% of it back towards Earth.

The other two important greenhouse gases are ozone and halocarbons. Despite the fact that  most of the greenhouse gases occur naturally in the atmosphere, some are man-made and the most well-known of these are fluorocarbons. Since the industrial revolution, human activities have also resulted in an increase in natural greenhouse gases, especially carbon dioxide. An increase in these gases in the atmosphere enhances the atmosphere's ability to trap heat, which leads to an increase in the average surface temperature of the Earth.

For a really strong greenhouse effect, we should look at Venus. Venus is similar to Earth in terms of size and mass, but its surface temperature is about 760~K. This is hot enough to melt lead! The Venusian atmosphere is mainly made up of carbon dioxide, a greenhouse gas. 

In summary, the habitable zone is the orbital range around a star within which surface liquid water could be sustained. Since water is essential for life as we know it, the search for biosignature gases naturally focuses on planets located in the habitable zone of their host stars. The habitable zone of the solar system looks like a ring around the Sun. Rocky planets with an orbit within this ring may have liquid water to support life. The habitable zone around a single star looks similar to the habitable zone in our Solar System. The only difference is the size of the ring. If the star is bigger than the Sun it has a wider zone, if the star is smaller it has a narrower zone.  It might seem that the bigger the star the better. However, the biggest stars have relatively short lifespans, so the life around them probably would not have enough time to evolve~\cite{Loeb:2016vdd}. The habitable zones of small stars face a different problem. Besides being narrow they are relatively close to the star. A hypothetical planet in such a region would be tidally locked~\cite{Seager}. That means that one half of it would always face the star and be extremely hot, while the opposite side would always be facing away and freezing. Such conditions are not very favorable for life.

The frequency $\eta_\oplus$ of terrestrial planets in and the habitable zone of solar-type stars can be determined using data from the {\it Kepler} mission~\cite{Borucki:2010zz,Borucki:2011nn,Batalha:2012gh}. Current estimates suggest $0.15^{+0.13}_{-0.06} < \eta_\oplus < 0.61^{+0.07}_{-0.15}$~\cite{Dressing:2013mid,Kopparapu:2013xpa}.

\section{Life outside the Habitable Zone}
\label{sec:3}

It has been pointed out that life outside the habitable zone may be possible on planets with subsurface oceans~\cite{Lingam:2017}.
Allowing for the possibility of subsurface ocean worlds yields a frequency of planets $\eta \sim 1$.
Now, because the  planets with subsurface oceans outside the habitable zone are more common than rocky planets in the habitable zone, one may wonder 
why do we find ourselves on the latter. The answer to this question most likely stems from the fact that ``we'' refers to an intelligent, conscious and technologically sophisticated species. In other words, albeit the probability of life on subsurface worlds may be non-negligible, it is quite plausible that  the likelihood of technological life could instead be selectively lowered. In this section we explore the possibility that life elsewhere could  have a non-carbon chemical foundation; e.g., in the spirit of~\cite{Reynolds,Wells}  we envision a race of intelligent silicon-based life forms.

Considering that Earth is the only reference point we have when studying life, it is unsurprising that biochemistry has always been connected to the elements of carbon, hydrogen, oxygen, and nitrogen. Moreover, carbon can form bonds with many other non-metals, as well as large polymers. These unique qualities have led many to argue that carbon is a pre-requisite for the existence of even very simple life. However, this must not necessarily be true. It has long been suspected that silicon and germanium can enter into some of the same kind chemical reactions than carbon does~\cite{Sagan}. Recent research in both chemistry and astrobiology has shown that it is theoretically quite feasible for silicon to form complex, self-replicating systems similar to the ones that produced the first, simple forms of life on Earth~\cite{Bains,Schulze-Makuch}. More concretely:
\begin{itemize}[noitemsep,topsep=0pt]
\item Silicon is able to form stable covalent bonds with itself, as well as stable compounds with carbon and oxygen~\cite{West}. These structures can form many diverse systems, including ring systems, which could be analogs to sugars, a key component of biochemistry on Earth. This stability is a prerequisite for  building  the complex chemical structures that support life on Earth, making silicon a strong contender. 
\item Silanols, the silicon containing analogues of alcohols have surprising solubility properties, with diisobutylsilane diol being soluble in water and hexane~\cite{Lickiss}. Solubility is another crucial factor in the development of life, since having a solvent and a substance is the model for early development of life that we see on Earth. 
\item Silicon's {\it chiral} properties. All life on Earth is made of molecules that twist in the same direction, that is they have an inherent {\it handedness}. In other words, each of life's molecular building blocks (amino acids and sugars) has a twin: not an identical one, but a mirror image. On Earth, the amino acids characteristic of life are all ``left-handed'' in shape, and cannot be exchanged for their right-handed doppelg\"anger. Meanwhile, all sugars characteristic of life on Earth are ``right-handed.'' The opposite hands for both amino acids and sugars exist in the universe, but they just are not utilized by any known biological life form. (Some bacteria can actually convert right-handed amino acids into the left-handed version, but they cannot use the right-handed ones as is.) This phenomenon of biological shape selection is called {\it chirality} -- from the Greek for handedness. We say that both sugars and amino acids on Earth are homochiral: one-handed.

 Though we are still unsure why it is that the molecules of carbon-based life choose only one orientation, it seems reasonable to require that in order for silicon to replicate the processes that originated life on Earth the molecules must also be chiral, and exist in a left- or right-handed forms in potential living environments. There is certainly reason to be optimistic: an observation of chirality in noncrystalline silica chiral nano-ribbons has been reported in~\cite{Okazaki}. 
\item Silicon's high reactivity is a barrier to forming complex structures on Earth, as this high rate of reaction leaves little time for construction. However, this only holds true for environments with a climate similar to earth. On the outskirts of the solar system, where the reactivity of carbon is severely impacted by the drop in temperature, silicon's high reaction rate could be the key to the development of life in these cryogenic environments, allowing it to flourish where carbon based life would be impossible. 

One probable environment for silicon life is liquid nitrogen~\cite{Bains}. Nitrogen is one of the few substances that can still dissolve silicon at very cold temperatures, as solubi. Additionally, silicon is able to form stable covalent bonds with nitrogen, as well as with itself. 
\end{itemize}

The habitable zone for silicon life would then depend on the area around a star in which nitrogen is a liquid. Neptune's moon, Triton, has been considered a candidate for surface level nitrogen lakes~\cite{Broadfoot}. Triton is the only large satellite in the solar system to circle a planet in a retrograde direction, i.e. in a direction opposite to the rotation of the planet. The retrograde orbit and Triton's relatively high density suggest that this satellite may have been captured by Neptune as it traveled through space several billion years ago. If this were the case, tidal heating could have melted Triton in its originally eccentric orbit, and the satellite might have been liquid for as long as one billion years after its capture by Neptune. However, presently Triton is quite cold, with a surface temperature of  38~K, and an extremely thin atmosphere (the atmospheric pressure at Triton's surface is about 14 microbars, 1/70,000th the surface pressure on Earth). Nitrogen ice particles might form thin clouds a few kilometers above the surface. Hence, even though the surface temperature is below the freezing point of liquid nitrogen
it is reasonable to assume that  the albedo of a hypothetical planet that could support silicon life will be similar to that of Triton,  $\alpha_{\rm Triton} \sim 0.6$~\cite{McEwen}.

Next, using (\ref{once}) we determine the habitable zone of silicon-based life for a main sequence star like our sun, with temperature
$T_\odot$ and radius $R_\odot$. We take the planetary surface temperature in between the
boiling and freezing point  of liquid nitrogen,
$63.15 < T_p/{\rm K} < 77.36$. Plugging in these values in (\ref{once}) we find that for a main sequence star like our sun, the habitable zone of silicon-based life stretches from 1.24~billion km  to 1.85~billion km from the star. We can now estimate what planets within the solar system fall into the silicon habitable zone during all parts of their orbit. The two planets closest to the silicon habitable zone are Saturn and Uranus. Saturn has a perihelion of 1.35 billion km  and  aphelion of  1.51 billion km, meaning that it is within the proper distance range for silicon biochemistry. However, Saturn is mostly a gas planet, and thus unsuitable for supporting any life. Uranus, on the other hand, has a perihelion of 2.75 billion km and aphelion 3.00 billion km, making it too cold for surface lakes or oceans of nitrogen. This result is also in agreement with the commonly accepted surface temperature of Uranus, roughly 57~K~\cite{Lunine:1993fz}, which is below the freezing point of nitrogen.

We expand our focus to include ultra-cool stars, such as  TRAPPIST-1A, as they are the most common stars in the Milky Way, and thus their orbiting planets are representative of ``average'' star systems. More concretely, M-dwarfs like Proxima Centauri and TRAPPIST-1 are 10 times more abundant than the Sun~\cite{Chabrier:2003ki,Robles:2008ww} and have stellar lifetimes that are about 100 to 1000 times greater~\cite{Tarter:2006dy,Adams:1996xe, Loeb:2016vps}. Furthermore, exoplanets around these stars are easier to detect (the transit signals produced by Earth-sized planets are 80 times stronger than the signal produced by similar planets transiting a Sun-like star) and their atmospheres can be analyzed via transit spectroscopy, thus enabling the ready detection of biomarkers~\cite{Fujii}. However, various physical mechanisms could act in concert to suppress the likelihood of Earth-based life on M-dwarf exoplanets relative to their counterparts around solar-type stars~\cite{Lingam:2018dky}. Nevertheless, this may not be the case for  silicon-based alien life forms. Herein, we evaluate the TRAPPIST-1 system as representative of ultra cool stars, for which $T_\star = 2,511~{\rm K}$, and $R_\star = 84,179.7~{\rm km}$~\cite{VanGrootel}. We generalize (\ref{once}) substituting $T_\odot$ by $T_\star$ and $R_\odot$ by $R_\star$, to find that the habitable zone for ultra-cool dwarf stars encompasses a distance range between   $1.6$~million km to $3.0$~ million km from the planet's star, whereas for silicon-based life on nitrogen lakes the habitability circumstellar region spans the orbital range within $28$~million km and $42$~ million km. This seems to indicate  the frequency of planets hosting any form of life must be extended. As for subsurface ocean worlds, we may take $\eta \sim 1$ for intelligent, conscious and technologically sophisticated species.

\section{Upper Limit on $\bm{\langle \xi_{\rm biotec} \rangle}$}

\label{sec:4}

Next, in line with our stated plan, we derive an upper limit on $\langle \xi_{\rm biotec} \rangle$. 
The present day star formation rate in the Galaxy is estimated to be $\dot M_\star = 1.65 \pm 0.19~M_\odot~{\rm yr}^{-1}$~\cite{Licquia:2014rsa,Chomiuk:2011fc}. This estimate has been derived assuming the Kroupa initial mass function (IMF)~\cite{Kroupa:2002ky,Kroupa:2003jm}. The shape of this IMF is lognormal-like and exhibits a peak around $M/M_\odot \approx 0.4$~\cite{Lada}, suggesting there are roughly 2 stars per $M_\odot$. Altogether, this yields $\Gamma_\star \approx 3~{\rm yr}^{-1}$. Before proceeding, we pause to note that the average star formation rate in the Galaxy could be about 4 times the current rate~\cite{Kennicutt:2012ea}.  Now, only 10\% of these stars are appropriate for harboring habitable planets. This is because the mass of the star $M_\star < 1.1 M_\odot$ to be sufficiently long-lived (with main sequence lifetimes larger than 4.5~Gyr) and $M_\star > 0.7 M_\odot$ to possess circumstellar habitable zones outside the tidally locked region~\cite{Seager}. The production rate of habitable planets is then
\begin{equation}
\zeta \sim 0.045 \left(\frac{\Gamma_\star}{3~{\rm yr^{-1}}}\right)  \left(\frac{\eta}{0.15} \right)~{\rm yr}^{-1} \,,
\end{equation}
with $0.15 < \eta <1$ and $3 <\Gamma_\star/{\rm yr}^{-1} < 12$.

Now,  a habitable planet must survive and remain in a habitable zone to present day. 
The potential hazard of nearby gamma-ray bursts (GRBs) has been estimated elsewhere~\cite{Anchordoqui:2017eog}. 
Long GRBs are associated with supernova explosions. When nuclear fuel is exhausted at the center of a massive star, thermal pressure can no longer sustain gravity and the core collapses on itself. If this process leads to the formation of a rapidly spinning black hole, accreted matter can be funneled into a pair of powerful relativistic jets that drill their way through the outer layers of the dying star.
 If the jet is pointing towards Earth, its high-energy emission is seen as a GRB. 

 The luminosity of long GRBs -- the most powerful ones -- is so intense that they are observed about once a day from random directions in the sky. The physical conditions in the dissipation region produce a heavy flux of photons with energies above about 100~keV. If one GRB were to happen nearby, the intense flash of gamma rays illuminating the Earth for tens of seconds could severely damage the thin ozone layer that absorbs ultraviolet radiation from the Sun. Indeed a fluence of $100~{\rm kJ/m^2}$ would create a depletion of 91\% of this life-protecting layer on a timescale of a month, via a chain of chemical reactions in the atmosphere. This would be enough to cause a massive life-extinction event~\cite{Ruderman,Thorsett:1995us,Dar:1997he,Thomas:2004vj,Thomas:2005gpa,Melott:2003rs,Piran:2014wfa}.

The critical time $t_c$  for life to arise and evolve becomes the dominant uncertainty on estimating the probability of having at least one lethal GRB, with a critical fluency $100~{\rm kJ/m}^2$. Life has been evolving on Earth for close to 4~Gyr~\cite{Mojzsis,Dodd}, but complex life is well under 1~Gyr old, and intelligent life is only a Myr old at most. In what follows we adopt $t_c =1~{\rm Gyr}$ and 4~Gyr as critical time intervals for life evolution~\cite{Loeb:2016vps}. With this in mind, the probability for a GRB to destroy an entire alien race $/\!\!\!\zeta$ is estimated to be~\cite{Anchordoqui:2017eog}
\begin{equation}
  0.044 \alt /\!\!\!\zeta \alt 0.22 \, .
\end{equation}
To derive the upper limit on $\langle \xi_{\rm biotect} \rangle$ we must take as fiducial parameters those giving the possible lowest value of 
$\langle \zeta_{\rm astro} \rangle \sim \zeta  /\!\!\!\zeta$, i.e.
\begin{equation}
  \langle \zeta_{\rm astro} \rangle   \sim 2 \times 10^{-3} \left(\frac{\Gamma_\star}{3~{\rm yr}^{-1}}\right)  \left(\frac{\eta}{0.15} \right) \left(\frac{/\!\!\!\zeta}{0.044}\right)~{\rm yr}^{-1} \, .
  \label{zetaastro}
\end{equation}
  
Finally, to determine the upper bound on $\langle \xi_{\rm biotec} \rangle$ we must decide on the possible minimum $L_\tau$. Herein we consider $L_\tau > 0.3~{\rm Myr}$ such that $c L_\tau \gg$ propagation distances of Galactic scales ($\sim 10~{\rm kpc}$). This would provide enough time to receive electromagnetic (and/or high-energy neutrino~\cite{Learned:2008gr}) signals from any advanced civilization living in the Milky Way which is trying to communicate with us.

The non observation of artificial signals from beyond Earth prevents an estimate of the event rate. Our best estimate of the proportion of cases which have a signal (i.e. an event) is zero, but there will be uncertainty in this estimate. Just because we have not seen an event yet does not mean we will never see one. We need a confidence interval for this estimate.
If a corresponding hypothesis test is performed, the confidence level (CL) is the complement of the level of statistical significance, e.g, a 95\% confidence interval reflects a significance level of 0.05. Because the number of events observed is zero, we cannot use the usual standard error estimate for the confidence interval. Instead, we use a small sample confidence interval for the estimate, based on the exact probabilities of the Binomial distribution. In the  absence of background, $N < 3.09$ determines the 95\% confidence interval~\cite{Feldman:1997qc}. 

Models with associated parameters in the right-hand-side of (\ref{Dshort}) predicting more than 3.09 events are excluded at 95\% CL. Then substituting (\ref{zetaastro}) into (\ref{Dshort}) with $L_\tau \sim 0.3~{\rm Myr}$ we obtain
\begin{equation}
\langle \xi_{\rm biotec} \rangle < 5 \times 10^{-3} \left(\frac{3~{\rm yr}^{-1}}{\Gamma_\star}\right)  \left(\frac{0.15}{\eta} \right) \left(\frac{0.044}{/\!\!\!\zeta}\right)  
\end{equation}
at the 95\% CL.

\section{Conclusion}
\label{sec:5}

We have re-examined the likelihood for alien civilizations to develop communication technology on the basis of the general assumption that life elsewhere could have a non-carbon chemical foundation.
We derived a conservative upper bound on the average fraction of living intelligent species of Earth-like planets that develop communication technology: $\langle \xi_{\rm biotec} \rangle < 5 \times 10^{-3}$ at the 95\%~CL. The upper limit can be up to a factor of 6 more restrictive if we assume there may be non-carbon based living organisms.

The Breakthrough Listen Initiative, announced in July 2015 as a 10-year 100M USD program, is the most comprehensive effort in history to quantify the distribution of advanced, technologically capable life in the universe~\cite{Isaacson,Lipman,Gajjar:2019rvn}. The search for
extrasolar biomolecular building blocks and  molecular biosignatures of extinct extraterrestrial life will soon be extended to the lunar surface~\cite{Lingam:2019}. The ability to detect alien life may still be years or more away, but the quest is underway. 

 \acknowledgments{This work has been
    supported by the by the U.S. National Science Foundation (NSF
    Grant PHY-1620661) and the National Aeronautics and Space
    Administration (NASA Grant 80NSSC18K0464). }

\end{document}